\documentstyle[12pt,epsfig]{article}

\begin{document}

\begin{center}
{\Large The $X(3872)$ and the 3941 MeV peak in $\omega J/\Psi$ }

\vskip 5mm
{D.V.~Bugg},   \\
{Queen Mary, University of London, London E1\,4NS, UK}
\end {center}

\begin{abstract}
Belle data for the $\pi \pi$ mass spectrum in $X(3872) \to \pi ^+\pi
^-J/\Psi$ are well fitted by $\rho + J/\Psi$ with $J^{PC} = 1^{++}$, but
are poorly fitted by the $\pi \pi$ S-wave.
This rules out quantum numbers $1^{+-}$, $2^{--}$ and $3^{--}$.
Formulae for partial wave amplitudes are given for all likely
$J^{PC}$ and decay modes of the 3941 MeV peak which Belle
observe in $\omega + J/\Psi$.
Angular correlations involving up to five angles may allow a
spin-parity determination, even with quite low statistics.
Special attention is given to the case where $X(3872)$ is a cusp
or quasi-bound state with the same quantum numbers as the 3941 MeV peak.
\end{abstract}

PACS Categories: 11.80.Et, 13.20.Fe, 13.20.He, 14.40.Lb

\section {Introduction}

The Belle collaboration observes a narrow peak at
3872 MeV in $\pi ^+\pi ^-J/\Psi$; it has
a width $<2.3$ MeV with 95\% confidence [1].
This peak is confirmed by CDF\,II [2]
and by the D0[3] and Babar [4] collaborations.
There has been extensive discussion of possible quantum numbers
and expected charmonium levels [5-14].
The fact that X(3872) lies very close to the $\bar D^0D^{*0}$
threshold and is narrow has prompted suggestions that it
may be a quasi-bound state of $\bar
DD^* + \bar D^*D$ [15].
T\" ornqvist ascribes this possibility to long range
attraction due to $\pi$ exchange [16]. A possibility explored
here is that it may be a cusp due to de-excitation of $\bar DD^*$ to
other open channels [17].

A new result is that Belle also observes
an enhancement near threshold in $\omega J/\Psi $ at $3941 \pm
11$ MeV with a width $\Gamma = 92 \pm 24$ MeV [18]. As a shorthand,
this peak will be called $Y(3941)$.

This paper discusses three somewhat disparate topics related to
quantum numbers $J^{PC}$ of $X(3872)$ and $Y(3941)$.
Section 2 examines fits to the $\pi \pi$ mass spectrum in
$X(3872) \to \pi ^+\pi ^- J/\Psi$ for all likely values of
the orbital angular momentum $L$ between $J/\Psi$ and $\pi \pi$.
For $L = 0$, the $\rho (770)$ gives an excellent fit, as Belle
remarked in
their first publication [1]; this supports $J^{PC} = 1^{++}$.
Fits with the $\pi \pi$ S-wave require discussion of how to parametrise
it; even the most conservative parametrisation gives
distinctly poorer fits because S-wave structure is broader
than $\rho (770)$. Discrimination against quantum numbers
$1^{+-}$ ($^1P_1$), $2^{--}$ ($^3D_2$) and $3^{--}$ ($^3D_3$)
is quite strong.

Section 3 reviews remaining possibilities.
The $Y(3941)$ lies close to predicted
radial excitations of $\chi _{c0}$, $\chi _{c1}$ and $\chi _{c2}$.
However, the observed decay width of $92 \pm 24$ MeV
is a factor $\sim 10$ larger than expected for $\omega J/\Psi$ alone.
The remaining width is likely to come from decays to $\bar D
D^*$ or $\bar D D$. Predictions of Eichten et al. [14] then favour
$J^{PC} = 1^{++}$. One is then faced with the possibility that
$X(3872)$ and $Y(3941)$ may both have $J^{PC} = 1^{++}$.
This can arise if $X(3872)$ is a cusp at the $\bar D D^*$ threshold
or a `molecular' state.
In the latter case, mixing between $1^{++}$ S and D-waves needs
experimental study.

Progress depends on angular momentum analyses of both
$X(3872)$ and $Y(3941)$.
Section 4 is intended to assist such analyses by providing tensor
formulae for partial wave amplitudes.
Up to now, discussions have depended on decays of $X$ or $Y$.
However, there is additional information available from the
production process $B \to K + X$ (or $Y$).
The full formulae involve correlations between five angles
describing
(i) this production process,
(ii) decay angles of $X$ (or $Y$) $\to C + D$ and
(iii) decay angles of $C$ and $D$.
Tensor expressions have the virtue of being written compactly
in Cartesian coordinates and can be programmed in a few lines
of code.
These formulae extend naturally to $\bar D D^*$ decays and indeed
most decays of the $B$ meson and charmonium states; only minor
substitutions of variables are required.
Hopefully these formulae may find application to a large
range of processes.
An Appendix discusses the
three-dimensional geometry of the $J/\Psi$ decay and the effect of the
Lorentz transformations to the rest frames of $X$ or $Y$.

Earlier, Pakvasa and Suzuki have given angular distributions for some
decays of $X(3872)$ [19]. Rosner [20] gives decay angular distributions
for important cases: $J^P = 0^+$, $1^+$ and $2^+$ S-wave decays to
$(\rho ^0$ or $\omega)+J/\Psi$ and for $J^P = 0^-$; he does not discuss
correlations between decays and the production process. Full formulae
for both production and decay are given here for all $J^P$ up to $3^-$.

Section 5 summarises conclusions and makes some suggestions about
the way angular momentum analyses might proceed.

\section {The $\pi ^+\pi ^-$ mass spectrum of $X(3872)$ decays}
Belle point out that this mass spectrum peaks at
the highest available mass, close to $\rho (770)$.
A second possibility which needs quantitative discussion is
interpretation in terms of the $\pi \pi$ S-wave.
This is a tricky point.

Fig. 1(a) shows the Breit-Wigner line-shape for $\rho (770)$ and
the intensity for $\pi \pi$ S-wave elastic scattering, denoted
by $S$.
Both intensities have peaks close to 770 MeV, but the S-wave
intensity is much broader.
This figure does not include the phase space
available for  $X \to \pi \pi J/\Psi$.

\begin{figure} [htb]
\begin{center}
\epsfig{file=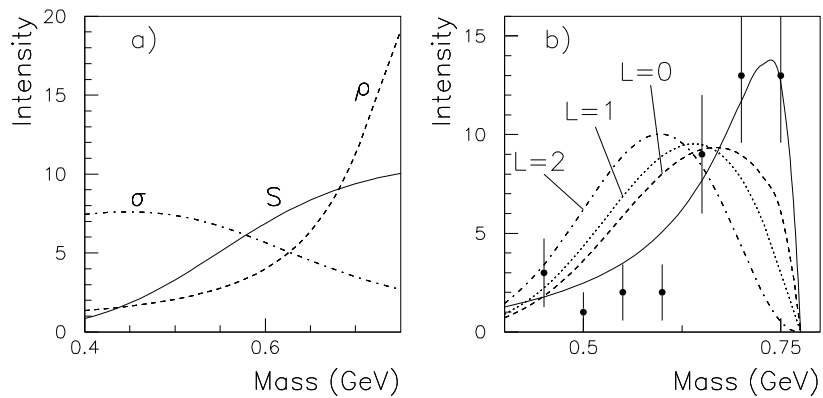,width=12cm}\
\vskip -6cm
\caption{(a) Line-shapes of $\rho(770)$,
$\pi \pi$ elastic scattering $S$, and the $\sigma$
pole $\sigma$.
(b) The $\pi \pi$ mass spectrum from Belle data on $X(3872) \to \pi
^+\pi ^-J/\Psi$, compared with the expected $\rho$ mass distribution
(full curve), and decays to $S$ with $L = 0$, 1 and 2.}
\end{center}
\end{figure}

In BES data for $J/\Psi \to \omega \pi ^+\pi ^-$,
a very different $\pi\pi$ peak is
observed at $\sim 500$ MeV  and is fitted with a $\sigma$ pole
at 541 MeV [21];
the fitted intensity is illustrated by the
chain curve labelled $\sigma$ in Fig. 1(a).
Data of E791 on $D^+ \to \pi ^+(\pi ^+ \pi ^-)$ are
similar but give a somewhat lower mass, 478 MeV and a narrower peak
[22]; for present purposes, the difference between these two is not
important.
What needs discussion is
the difference between the $\sigma$ and $S$ peaks.
What should be expected in $X(3872) \to [\pi \pi J/\Psi ]$?

Actually both $\sigma$ and $S$ curves may be fitted by
a single formula, given in Ref. [23] (with a correction of a
sign error in the Erratum).
The reaction $J/\Psi \to \omega \pi \pi$ is a `hard' process
producing the $\pi \pi$ pair with a large momentum
transfer. The Breit-Wigner denominator alone creates the peak.
However, $\pi \pi$ elastic scattering is a `soft' process and
the numerator of the amplitude contains an additional zero
at $s = 0.5 m^2_\pi$, just below
threshold. This is the Adler zero of Chiral Perturbation Theory. In
elastic scattering, the side of the pole close to $s = 0$ is sheared
away by the Adler zero. For elastic $\pi \pi$ scattering, the intensity
is then rather featureless at low mass and peaks only at 800 MeV. For
complex $s$, the singularity resembles a mountain ridge running at
$45^\circ$ to the real $s$ axis. The peak moves dramatically with
$\mathrm {Im}\, s$; the resonance is at 925 MeV for real $s$, but at
540 MeV for $\mathrm {Im}\, s = 0.25$ GeV.

The decay $X(3872) \to S J/\Psi $ involves a small
momentum transfer to the $\pi\pi$ system,  so
the $\pi \pi$ elastic scattering amplitude $S$ is likely to be
more appropriate. However, it is not yet known how
rapidly the Adler zero disappears with increasing momentum
transfer.
The approach here is to fit using the $\pi \pi$ elastic
scattering amplitude; this is the most conservative approach,
closest to data on $X(3872)$.
Any effect of
the disappearing Adler zero will shift the peak downwards
in mass, away from the $\rho$, making fits to the $\pi \pi$
S-wave even less likely.

Fig. 1(b)  shows the $\pi \pi$ mass spectrum from Fig. 3 of
Ref. [1] in 50 MeV bins.
The full curve shows the expected signal for $L=0$
decays, folding in the available phase space for
$X(3872) \to \rho J/\Psi $. This gives an excellent fit with
$\chi ^2 = 4.6$ for 6 degrees of freedom.
The mass spectrum for  $[\rho J/\Psi ]_{L=1}$ gives
$\chi ^2 = 12.1$ and cannot be excluded; however, a
strong rate would be surprising, since the
$\rho$ is so close to the top of phase space.

The dashed, dotted and chain curves show fits to Belle data using curve
$S$ of Fig. 1(a) and $L=0$, 1 or 2.
The $L = 0$ fit is affected much more by the phase space cut-off than
the $\rho$ fit.
It gives a poor $\chi ^2$ of 19.4;
$J^{PC} = 1^{--}$ is highly unlikely if $X(3872)$ is a
charmonium state, which should appear in $e^+e^-$ interactions.
Seth [24] points out that $X(3872)$ could be a $J^{PC} = 1^{--}$
glueball; a pure glueball would not couple to $e^+e^-$.

For $L = 1$ and 2, the fits are
very poor with $\chi ^2 = 49 $ and 343 respectively. Blatt-Weisskopf
centrifugal barriers are included with a radius of $0.8$ fm. The fits
are much worse if the $\sigma$ pole of Fig. 1(a) is used with no Adler
zero.
For $[S J/\Psi ]_{L=1}$, isospin
conserving decays are $0^{+-}$ (exotic), $1^{+-}$ or $2^{+-}$
(exotic). For $J^{PC} = 1^{+-}$, Belle already find poor agreement with
the decay angular distribution [25]. Also Babar find no evidence for
$X(3872) \to \eta J/\Psi$ [26]; this makes $1^{+-}$ doubly unlikely.
For $L = 2$, the $J^{PC} = 2^{--}$ ($^3D_2$) and $3^{--}$
($^3D_3$) possibilities are definitely excluded.

Babar find no evidence for charged $X(3872)$ [27], so $\rho J/\Psi$
decays are probably isospin violating.
For $L=0$, all of $J^{PC} = 0^{++}$,
$1^{++}$ and $2^{++}$ are clear candidates.
The possibilities for $L=1$ decays are
$J^{PC} = 0^{-+}$, $1^{-+}$ (exotic), $2^{-+}$ and $3^{-+}$ (exotic);
although less likely, these are not excluded at present.
However, the $\eta _c''$ is expected to lie above 4000 MeV.

\section {General considerations}
Table 1 lists possible $J^{PC}$ for each decay;
those for $C = +1$ are candidates for both $X(3872)$ and $Y(3941)$,
while those for $C = -1$ refer only to $X(3872) \to S J/\Psi$.
The notation is that $s$ denotes combined spins of $J/\Psi$
with $\omega$ (or $\rho $); $L$ is the orbital angular
momentum in decay channels. $L=3$ decays are omitted as unlikely for
$X$ and $Y$ and to $\bar D D^*$.

\begin {table}[t]
\begin {center}
\begin {tabular} {cccccc}
\hline
$J^{PC}$  & $s,L$  & $\bar D D^*$ & $\bar DD$ & $S J/\Psi$ \\
          &        &    $s=1$   &   $s=0$ &  $s=1$   \\\hline
$0^{++}$ & $s=0,L=0$ & - & $L=0$ & -  \\
         & $(s=2,L=2)$ & & & & \\
$1^{++}$ & $s=1,L=0$ or 2 & $L=0$ or 2   & - & -  \\
         & $(s=2,L=2)$ & & & & \\
$2^{++}$ & $s=2,L=0$ (or 2) & $L=2$ & $L=2$ & - \\
         & $(s=0$ or
 1,$L=2)$ & & & & \\
$0^{-+}$ & $s=1,L=1$ & $L=1$ & - & -  \\
$2^{-+}$ & $s=1$ or 2,$L=1$ & $L=1$  & - & - \\
$1^{+-}$ &  - & $L=0$ (or 2) & - & $L=1$ \\
$2^{--}$ &  - & $L=1$  & - & $L=2$ \\
$3^{--}$ &  - & - & - & $L=2$ \\\hline
\end {tabular}
\caption{Possible $J^{PC}$ for $Y(3941) \to \omega J/\Psi $,
$X(3872) \to \rho J/\Psi$ ($C = +1)$ or $S J/\Psi $ $(C = -1)$;
also possible $\bar D D^*$ and $\bar DD$ decays.
Those in parentheses are less likely. }
\end {center}
\end {table}

Eichten et al. [14] survey charmonium states likely from 3800 to 4000
MeV and calculate decay widths. The lowest $^1D_2$, $^3D_2$ and $^3D_3$
are expected somewhat below $X(3872)$. The radial excitations of
$^3P_0$, $^3P_1$ and $^3P_2$ are predicted in the general mass range of
$Y(3941)$ and are serious candidates.

I wish to draw attention to the possibility that $X(3872)$ and
$Y(3941)$ may both have $J^{PC} = 1^{++}$. There are two tentative
pointers in this direction for $Y(3941)$, though
experimental proof is needed from partial wave analysis.

\subsection {The width of $Y(3941)$}
The observed total width of $Y(3941)$ is $92 \pm 24$ MeV.
This width cannot plausibly be explained by the $\omega J/\Psi$
channel alone.
If $Y(3941)$ is a charmonium state, the production of
a decay $\omega$ requires formation of an additional $\bar uu +
\bar dd$ pair via a process involving at least two gluons.
This should lead to a width of the same order as $\chi _{c0}$,
$\chi _{c1}$ and $\chi _{c2}$, i.e. 1 to 16 MeV.
The remaining width is likely to come from $\bar DD^*$ or
$\bar D D$ decays.

Eichten et al. [14] predict for the $^3P_1$ radial excitation
a $\bar DD^*$ width of 150 MeV at a mass of 3968 MeV;
if the width is proportional to $\bar DD^*$ phase space,
it scales to 127 MeV at 3941 MeV, not too far from
the observed value. For $J^{PC} = 0^{++}$ or $2^{++}$, they
calculate much smaller widths for $D\bar D$ decays, although there
is some sensitivity in these calculations to radial
wave functions.
For $J^{PC} = 0^{++}$, they find a 40 MeV width to
$D_s\bar D_s$ at 3968 MeV, but the threshold is at 3938
MeV, so this channel is unlikely to account for the observed
width of $Y(3941)$.

The agreement of the $Y(3941)$ width with the calculation
of Eichten et al. for $J^{PC} = 1^{++}$ is a mild pointer.
It is obviously important to search for decays to
$\bar DD^*$, $\bar DD$ and $\bar D_sD_s$.

\subsection {The production process}
In the Belle and Babar experiments, the $Y(3941)$ and
$X(3872)$ are produced via the $(V - A)$ current.
Allowed transitions are to $1^+$ and $1^-$, and to $0^+$
via the divergence of the axial current.
These are just the channels which have been observed in
$B$ decays.
A feature of Belle data is that there is a strong
$K\chi _{c1}(3510)$ signal in Fig. 3 of Ref. [25].
There is also a conspicuous $K\psi (3770)$ signal [28].
Both are reached by allowed V or A transitions.
The final state $K\chi _{c0}$ has been observed [29] and
also $KD_s(2317)$ [30].

However, it is difficult to estimate the strengths of forbidden
transitions.
The $W$ boson is absorbed into the final state
and the kaon is radiated with one unit of orbital angular
momentum (in addition to orbital angular momentum involved
within forbidden $V$ or $A$ interactions themselves).
The weak interaction is pointlike; angular momentum
transfers depend on differences in the radial wave functions
of $B$ and charmonium states.

The strength of the observed $X(3872)$ and $Y(3941)$ signals
will be taken here as a hint that they may both have $J^{PC} = 1^{++}$,
though
$0^{++}$ is also a serious possibility; $2^{++}$ is first forbidden.

\subsection {A bound state or cusp}
If the $X(3872)$ is a bound state of $\bar D D^*$, it should
appear at low masses in that channel, in the same way as the
threshold $NN$ $^3S_1$ amplitude is strong and can be related
to the deuteron pole.
If the effective range is close to the pion Compton wave-length,
the phase shift will drop through $90^\circ$ at $\sim 4$ MeV
above the $\bar D D^*$ threshold. One then expects a strong
threshold $\bar D D^*$ signal over a mass range of 5--10 MeV.

A possibility is that the $X(3872)$ is a threshold cusp.
The way a cusp works is explained in Ref. [17], but will be
repeated here briefly.
Suppose the low mass $\bar D D^*$ system de-excites to open
channels;
obvious possibilities are $[S \eta _c]_{L=1}$, $S \chi _{c0}$ and
the discovery channel $\pi \pi J/\Psi$.
Of these, the first is likely to be the strongest, since the available
momentum is well above the $L = 1$ centrifugal barrier.

Close to threshold, the cross section for $\bar D D^*$ de-excitation
to these open channels follows the familiar $1/v$ behaviour of thermal
neutron scattering.
The imaginary part of the $\bar D D^*$ elastic scattering
amplitude $f_S$ is given by $k/(4\pi)$ times the total cross
section.
The factor $k$ for centre of mass momentum cancels with the
$1/v$ factor above threshold, producing a constant amplitude
to first approximation.
However, there is a step in $a$ at threshold and hence in $\mathrm
{Im}\, f_S$. The real part of the amplitude is given by a
dispersion relation: \begin {equation} \mathrm {Re}\, f_S(s) = \frac
{1}{\pi }\, P \int \frac {\mathrm {Im}\, f_S(s')\, ds'}{s' - s}. \end
{equation} The step in $\mathrm {Im}\, f_S$ gives rise to a peak in
$\mathrm {Re}\, f_S$ at threshold.
Fig. 2 illustrates these effects from a model calculation whose
parameters will be explained shortly. The result is a narrow peak in
the intensity, Fig. 2(b).

\begin{figure}
\begin{center}
\epsfig{file=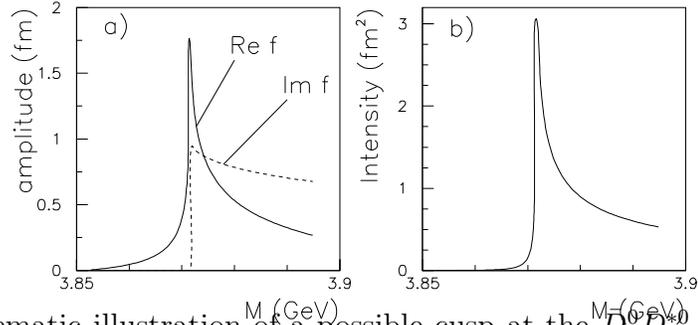,width=10cm}\
\vskip -55mm
\caption{Schematic illustration of a possible cusp
at the $\bar D^0 D^{*0}$ threshold: (a) Real and imaginary parts
of the elastic $\bar D^0 D^{*0}$ amplitude, (b) the corresponding
intensity. }
\end{center}
\end{figure}

This cusp appears not only in $\bar D D^*$ elastic scattering, but in
any channel coupled to the $\bar D D^*$ S-wave. The scattering
amplitude is an analytic function of $s$ with coupling constants
$g_i$ to individual channels $i$.
A familiar example is a Breit-Wigner resonance for coupling
of two channels 1 and 2:
$$
f_{12}(s) = \frac {g_1g_2}{M^2 - s - iM[g_1^2 \rho _1(s) + g^2_2 \rho
_2(s) + \ldots ]}.
$$
Even if there is no resonance, the $s$-dependence of $f_{12}$
is common to all channels coupled to the $\bar D D^*$ threshold
and the cusp appears in all of them.

At present, the strengths of couplings to open channls is not known.
There is competition between decays of $D^*$ to $D\pi$ and $D\gamma$
and  de-excitation to open channels.
There is also a question of whether long-range meson exchanges are
attractive or repulsive; repulsion would
shield the interaction close to threshold and eliminate the cusp.

The illustrative calculation of Fig. 2 is made along the
lines of Ref. [17] using a
scattering length approximation $k\cot \delta = 1/a$, where $a$ is
complex.
If the S-wave amplitude is written as $f_S = (e^{2i\delta } -
1)/2ik$, the S-wave amplitude near threshold is \begin {eqnarray} f_S
&=& \frac  {a}{1 - iak} = \frac {a + i|a|^2k}{1 + 2k\, \mathrm {Im}\, a
+ k^2|a|^2}, \\ |f_S|^2 &=& \frac {|a|^2}{1 + 2k\, \mathrm {Im}\, a
+k^2|a|^2}.
\end {eqnarray}

Ref. [17] makes comparisons with $\bar pp$
data, where the fitted scattering length is $\mathrm {Im}\, a = 1.8$ fm.
For $\bar D D^*$, $\mathrm {Im}\, a$ will be smaller, since open
channels are much weaker. If the $Y(3941)$ is a simple Breit-Wigner
$1^{++}$ resonance with a width proportional to $\bar D D^*$ phase
space, the scattering length is 0.4 fm and the effective range $-1.5$
fm. However, the imaginary part of the scattering length is a matter of
guesswork; it need not be related to the $Y(3941)$ resonance. Fig. 2
illustrates the cusp in $\bar D^0 D^{*0}$ elastic scattering for a
scattering length taken to be $0.4 +i 0.4$ fm. The step in the
real part represents an effective attraction,
which helps bind any `molecular' state such as  that suggested by
Braaten and T\" ornqvist. If the net effect of this attraction and
meson exchanges is strong enough, a bound state appears.

Experimentally, the $X(3872)$ peak is at the $\bar D^0 D^*$
threshold, not the threshold for charged particles.
This charge dependence can arise from a combination of
$\pi$, $\rho$ and $A_1$ exchanges.
A large isospin violation is then inevitable, since the $\bar D D^*$
phase shift changes by $90^\circ$ over $\sim 4$ MeV; this is
only half the mass difference between the neutral and charged
thresholds. Isospin violation is consistent with the observed decays to
$\rho J/\Psi$. For this reason, the second cusp at the $\bar D^-
D^{*+}$ threshold is not shown on Fig. 2, though it could in principle
be present. The height of the cusp
depends on the actual scattering length; the cusp gets wider as
$\mathrm {Im}\, a$ gets smaller.

A scenario worth consideration is that
there is a regular charmonium $1^{++}$ state at 3941 MeV,
and in addition a threshold cusp or `molecular' state
at the $\bar D D^*$ threshold.
If $\pi$ exchange plays a strong role, as T\" ornqvist
suggests [16], it will produce a tensor interaction
which mixes S and D-waves, as in the deuteron.
In the next Section, amplitudes will be given for the
D-wave possibility.
That mixing could well be absent for a pure cusp,
and might allow a distinction between a cusp and a
bound state. An alternative is that the $1^{++}$ state
is pulled down to the $\bar D D^*$ threshold and
$Y(3941)$ has $J^{PC} = 0^{++}$ or $2^{++}$.

\section {Angular Momentum Algebra}
Amplitudes will be written for $B \to K+Y(3941)$,
abbreviated henceforth as $Y$; those for $Y \to \bar D D^*$ and $X(3872)
\to S J/\Psi$ may be obtained by simple interchanges of symbols.
These expressions are easily carried over to a wide variety of
decays of the $B$ meson and charmonium for $s$ and $L \le
2$, $J$ up to 3. Tensor expressions will be used,
following the methods of Zemach [31]. However, it is convenient to
simplify the algebra from the outset by working in the rest frame of
$Y$. In this system, orbital angular momentum $\ell$ in the production
process is described by a 3-vector $K_\mu$ for the kaon momentum. The
orbital angular momentum between $J/\Psi$ and $\rho$ (or $\omega $) is
likewise constructed from the 3-vector $J_\mu$ for the $J/\Psi$
momentum. This greatly simplifies the algebra.

While this paper was being written, Rosner [20] produced a
paper giving expressions for angular dependence of decays.
He does not include correlations with the production
process, which give additional information included
in the tensor amplitudes written here.
His expressions for decay distributions have been checked against
formulae given here.
As an aid to experimentalists, amplitudes will be related
to the axes he chooses.

\subsection {Choice of axes}
Fig. 3 sketches vectors and angles
for the case Rosner considers: $X \to \rho J/\Psi$.
A key point is that polarisation $\vec e$ of the $J/\Psi$ is
orthogonal to the
axis of decays to  $e^+e^-$ (or $\mu ^+\mu ^-)$ in its rest
frame.  The leptons are shown
as they appear in this frame.
This lepton axis is adopted as my $z$-axis, but
amplitudes are written in the rest frame of $J/\Psi$ and $\rho$ (or
$\omega )$.
The choice of the $xy$ plane around the $z$ direction is not important.

\begin{figure}
\begin{center}
\epsfig{file=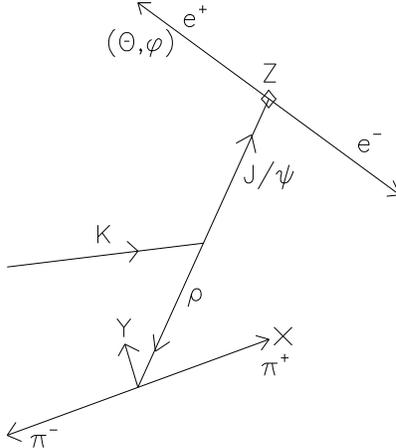,width=8cm}\
\vskip -1.5cm
\caption{Illustration of kinematics; $\theta$ is the polar angle
of the $e^+$ and $\phi$ its azimuthal angle in XYZ axes.}
\end{center}
\end{figure}

Rosner chooses his $X$ axis along the $\rho$ direction in Fig. 3
and defines the $XY$ plane to contain the decay pions.
His $Z$-axis is then normal to the $\rho J/\Psi$ axis.
He describes the $e^+e^-$ axis by angles $\theta, \phi$.
The Appendix gives the components  of
$J/\Psi$ polarisation in his axes, eqn. (26).
It also gives the effect of the Lorentz boost from
the $J/\Psi$ rest frame to that of $Y$ or $X$, eqn. (27).
This result for $\vec e$ is needed in equations below. The
Lorentz transformation has a small effect, so $\vec e$ remains
approximately transverse in that system.

The spin 1 of the $\rho$ is described by the 4-vector
\begin {equation}
Q = k(\pi _1) - k(\pi _2) - \frac {M^2_1 - M^2_2}{M^2_\rho }
[k(\pi _1) + k(\pi _2)].
\end {equation}
Ignoring the small mass difference between charged and
neutral pions, this becomes the 4-vector $[k_1 - k_2]^\mu $.
The time-component $E_1 - E_2$ is small but easily included
into the calculations.
Expressions for $\omega$ decay are likewise
described by a 4-vector $W_\mu$.
Amplitudes will be written here for $\omega J/\Psi$ decays;
those for $\rho J/\Psi $ decays are obtained
by substituting $Q$ for $W$.

\subsection {$\omega $ decay}
The $\omega$ decay may be viewed as $\omega \to [\rho \pi ]_{L=1}$.
For $\omega \to \rho ^+\pi ^-$:
\begin {equation}
W_\mu = \epsilon _{\mu \alpha \beta \gamma}
(k_+ - k_0)^\alpha
k_-^\beta (k_+ + k_- + k_0)^\gamma
= 2\epsilon _{\mu \alpha \beta \gamma} k_+^\alpha
k_-^\beta k_0^\gamma;
\end {equation}
the latter follows from energy-momentum conservation and
properties of $\epsilon _{\mu \alpha \beta \gamma}$.
The tensor
$\epsilon _{\mu \alpha \beta \gamma}$
has unit elements and is fully anti-symmetric:
if $\epsilon _{1234} = 1$, other elements are obtained by
changing the sign when adjacent elements are interchanged:
e.g. $\epsilon _{1243} = -1$.
All elements with two identical indices are 0.

To take account of the Breit-Wigner amplitude for the
$\rho$ and centrifugal barriers, $W_\mu$ should be multiplied by
\begin {equation}
f_\omega  =
\frac {B_1(k^{+0})B_1(k^-)}{(M^2 - s_{+0} - iM\Gamma )}
+ \frac {B_1(k^{0-})B_1(k^+)}{(M^2 - s_{0-} - iM\Gamma
)} + \frac {B_1(k^{-+})B_1(k^0)}{(M^2 - s_{-+} - iM\Gamma )}.
\end {equation}
In the first term, $B_1(k^-)$ is the $L=1$
centrifugal barrier factor
\begin {equation} B_1(k^-) = 1/[k^2_- + 0.03894/R^2(fm)]^{1/2}
\end {equation}
where $k_-$ is the momentum in GeV/c of the $\pi ^-$ in the
$\omega$ rest frame; $R \simeq 0.8$ fm.
Also $k^{+0}$ denotes the momentum of either $\pi$ in the
$\rho ^{+0}$ rest system.
The factor $f_\omega$ produces a mild modulation of the intensity
of $\omega \to \pi ^+\pi ^- \pi ^0$ decays over its Dalitz plot,
but has no effect on angular correlations.

In the $\omega$ rest frame, $(k_+ + k_- + k_0)^\gamma$ has only a
time component, so $W_\mu$ reduces to
$\epsilon _{\mu \alpha \beta 4}k^\alpha_+k^\beta_-M$.
The factor $M$ for $\omega$ mass is not essential; remaining
factors reduce to the 3-dimensional vector product $\epsilon _{\mu \alpha
\beta}k^\alpha _+k^\beta _- = k_+\wedge k_-$; so $W_\mu$ is along the
normal to the $\omega$ decay plane. It gives a
weight factor $|k_+k_-|f_\omega$. In the $Y$ rest frame, $W_\mu$
from eqn. (5) does
have a small time-component.

\subsection {$1^{++} \to [\omega J/\Psi ]_{s=1,L=0}$}
The amplitude for this process in the $Y$ rest frame is
\begin {equation}
v_\mu = \epsilon _{\mu \alpha \beta 4}
e^\alpha W^\beta M^4 BW(Y)
\equiv \epsilon _{\mu \alpha \beta } e^\alpha W^\beta  BW(Y);
\end {equation}
form factors for the decay may have more effect than the
variation of $M$ over the resonance, so it can be dropped;
$BW(Y)$ is the Breit-Wigner amplitude for $Y(3941)$.
Later expressions will omit the factor $BW$ and centrifugal barrier
factors, which are implied.
In principle some centrifugal barrier
factor is required for $\ell = 1$ in the production process.
However, this form factor is likely to vary negligibly over
the width of $Y(3941)$ and can be ignored.

The full matrix element for $B \to 1^{++}$, $1^{++} \to
[\omega J/\Psi ]_{L=0}$ is then the scalar product
$\vec K.\vec e\wedge \vec W = \vec e.\vec W \wedge \vec K$.
Here $\vec e$ has only $x$ and $y$ components in my axes,
say $(\cos R, \sin R)$.
The intensity is given by the average over $R$, using
$<\cos ^2 R > = <\sin ^2 R> = 1/2$.
In order to avoid this integration, a simple trick may be
used in computer programmes.
The $e_x$ and $e_y$ components may be replaced by 1 and
$i = \sqrt {-1}$.
Then intensities are obtained by taking the modulus
squared of expressions like $v_\alpha$.

\subsection {$0^{++} \to [\omega J/\Psi ]_{s=0,L=0}$}
For the production process, $\ell = 0$, so the matrix element
is simply the scalar product
$M = e^\alpha W_\alpha$.
One should form the 4-vector product,
though both $e_4$ and $W_4$ are numerically small:
\begin {equation}
P = e_1W^1 + e_2W^2 + e_3W^3 - e_4W^4.
\end {equation}

\subsection {$2^{++} \to [\omega J/\Psi ]_{s=2,L=0}$}
The fully relativistic expressions for tensors are
given in Ref. [32]. Here, they are simplified to the
rest frame of $Y$.
For spin 2, one needs
the tensor
\begin {equation}
T_{\alpha \beta} = e_\alpha W_\beta + e_\beta W_\alpha - {2\over 3}
\delta _{\alpha \beta }(e_\mu W^\mu ),
\end {equation}
where $\alpha$, $\beta$ and $\mu$ run from 1 to 3.
This last term eliminates the scalar term which would
otherwise be mixed into $T_{\alpha \beta}$;
it makes the tensor traceless.

The $\ell  = 2$ operator for the production process is likewise
constructed as
\begin {equation}
\tau_{\alpha \beta} = K_\alpha K_\beta  - {1\over 3}\delta _{\alpha
\beta } (K^2).
\end {equation}
As a reminder, $K$ is the 3-momentum of the kaon in the rest
frame of $Y$.
The required matrix element for production and decay is then
$M = \tau _{\alpha \beta }T^{\alpha \beta }$.
There is valuable information contained in the dependence on
production angles through $\tau _{\alpha \beta }$.
However, after integrating over all $K$ directions, one finds the
same decay angular distributions as given by Rosner's eqn. (17).
This result arises from the fact that $d\sigma /d\Omega = M^*M$
contains a term $\tau _{\alpha \beta }\tau ^{\beta '\alpha '}$.
Integrating over angles, this term becomes
$\delta _{\alpha \alpha '}\delta _{\beta \beta '}$ for all
amplitudes.

\subsection {Further amplitudes}
After these preliminaries, it is easy
to write down remaining expressions.
That for $1^{++} \to [\omega J/\Psi ]_{s=1,L=2}$ is
\begin {equation}
M(1^{++} \to [\omega J/\Psi ]_{s=1,L=2}) = K^\alpha \tau '_{\alpha
\beta } v^\beta,
\end {equation}
where $v^\beta$ is given by eqn. (8) and $\tau '$ is the $L=2$ operator
\begin {equation}
\tau '_{\alpha \beta } = J_\alpha J_\beta - {1\over 3}
\delta _{\alpha \beta }(J^2);
\end {equation}
as a reminder, $J$ is the 3-momentum of the $J/\Psi$ in the
rest frame of $Y$.
Here the centrifugal barrier $B_2$ for $L = 2$  is needed:
\begin {eqnarray}
B_2 &=& 1/[(J^2 + Z)J^2 + Z^2]^{1/2}, \\\nonumber
Z &=& 0.11682/R^2(fm).
\end {eqnarray}

Using both this and the $L=0$ amplitude eqn. (8) in the fit to $Y(3941)$
adds freedom which will need to be carefully controlled.
The signature for the $L=2$ amplitude is an interference with
the S-wave proportional to $J^2$. If the $X(3872)$ and
$Y(3941)$ have the same quantum numbers, they will have
orthogonal wave functions, hence orthogonal amplitudes
to a single channel, e.g. $\bar D D^*$ or $\omega J/\Psi$.

The amplitude for $1^{++}$
decaying to $s = 2$, $L = 2$ is less likely, since $s = 2$ does not mix
with $s = 1,L=0$. It gives the same decay angular distribution after
integrating over $K$ and $J$, but may be ambiguous with $s=1,L=1$ for
low statistics. It is the prototype for combining spins 2 and 2 to make
$J = 1$:
\begin {eqnarray} M(1^{++} \to [\omega J/\Psi
]_{s=2,L=2}) &=& K^\alpha V _{\alpha } \\\nonumber V _\alpha  &=&
\epsilon _{\alpha \beta \gamma 4}(T^{\beta \mu } {\tau '} _\mu ^\gamma
- T^{\gamma \mu }{\tau '} _\mu ^\beta )M^4 \\
     &\equiv & 2 \epsilon _{\alpha \beta \gamma }T^{\beta \mu }
{\tau '} _\mu ^\gamma .
\end {eqnarray}
For $J^{PC} = 2^{-+}$, there are two amplitudes with $s = 1$ and 2:
\begin {eqnarray}
M(2^{-+} \to [\omega J/\Psi ]_{s=1,L=1}) &=& \tau
^{\alpha \beta } [v_\alpha J_\beta + v_\beta J_\alpha - {2\over 3}
\delta _{\alpha \beta }(\vec
v.\vec J)], \\
M(2^{-+} \to [\omega J/\Psi ]_{s=2,L=1}) &=&
\tau ^{\alpha \beta }V^2_{\alpha \beta } \\\nonumber
V^2_{\alpha \beta } &=& \epsilon _{\alpha \mu \nu 4}T_\beta ^\mu
J^\nu M^4 + \epsilon _{\beta \mu \nu 4}T^\mu _\alpha J^\nu M^4  \\
               &\equiv & (\epsilon _{\alpha \mu \nu }T_\beta ^\mu
  + \epsilon _{\beta \mu \nu }T^\mu _\alpha) J^\nu .
\end {eqnarray}
This is the prototype for combining spins 2 and 1 to make 2.

The amplitude for $0^{-+}$ is given by the simple contraction
\begin {equation}
M(0^{-+} \to [\omega J/\Psi ]_{s=1,L=1}) = J^\alpha v_\alpha.
\end {equation}

For $J^{PC} = 2^{++}$, there are three $L = 2$ amplitudes with
$s = 0$, 1 and 2.
The first with $s = 0$ is given by the simple tensor contraction
$\tau ^{\alpha \beta }\tau '_{\alpha \beta }$.
The second for $s = 1$ is obtained from eqn. (19) by replacing $J$ by
$v$ and $T$ by $\tau ' $. The third is
\begin {eqnarray}
M(2^{++} \to [\omega J/\Psi ]_{s=2,L=2}) &=& \tau
^{\alpha \beta } Y_{\alpha \beta }  \\
Y _{\alpha \beta } &=& T_\alpha ^\mu
\tau ' _{\mu \beta } + T_\beta ^\mu \tau ' _{\mu \alpha }
- {2\over 3} \delta_{\alpha \beta }(T^{\mu \nu }\tau '_{\mu
\nu }).
\end {eqnarray}
There is little reason to expect these
amplitudes to compete with $L=0$. All four have the same dependence on
the angle between lepton and $\vec W$, and this will hopefully separate
them from other spin-parities. The separation of possible $s = 0$, 1
and 2 amplitudes with $L=2$ may be only of academic interest.

The $3^{--} \to [S J/\Psi ]_{L=2}$ amplitude constructed from $s=1$,
$L = 2$ is
\begin{equation}
V^3_{\mu \nu \lambda } = \tau ' {_{\mu \nu
}}e_\lambda + \tau ' {_{\mu \lambda }}e_\nu
 + \tau ' {_{\nu \lambda }}e_\mu - {2\over 5}(\tau ' {_{\mu
 \alpha}}\delta _{ \nu \lambda } + \tau ' {_{\nu \alpha}}\delta_{\lambda
 \mu} + \tau ' {_{\lambda \alpha}}\delta _{\mu \nu })e^\alpha.
\end {equation}
To form the full amplitude, this needs to be contracted with the
operator for production with $\ell = 3$:
\begin{equation}
\tau _{\mu \nu \lambda }=K_\alpha K_\beta K_\gamma - {1\over 5}(K^2)
(K_{\mu }\delta _{\nu \lambda} + K_{\nu }\delta _{\mu \lambda} +
K_{\lambda }\delta _{\mu \nu}).
\end {equation}

\subsection {Other channels}
Amplitudes for $S J/\Psi$ may be constructed along the same
lines. For $J^{PC} = 1^{+-}$, one needs to combine $\vec e$ with $
\vec J$
to make total spin 1, as in eqn. (8), replacing $W$ by $J$ and
replacing $\rho$ by the $S$ amplitude.
For $J^{PC} = 2^{--}$ one combines s = 1 with L = 2, in analogy to
eqn. (19), replacing $J$ by $e$ and $T$ by $\tau '$.
Amplitudes for $1^{--} \to S J/\Psi $ with $L = 0$ and 2 are:
\begin {eqnarray}
M(1^{--} &\to & [S J/\Psi ]_{L=0}) = K^\alpha e_\alpha \\
M(1^{--} &\to & [S J/\Psi ]_{L=2} ) = K^\alpha \tau ' {_{\alpha \beta }}
e^\beta.
\end {eqnarray}
The decay $B \to K+J/\Psi$ is also given by eqn. (25).
Production of $\Psi (3770)$ with decay to $[S J/\Psi ]_{L=2}$
is given by eqn. (26);
for decays to the $\pi \pi$ D-wave, $M = K^\alpha T_{\alpha \beta
} e^\beta$.
There is some evidence for $\psi (2S)$ decays to $\phi \pi ^0$
and
$\omega \eta $ [33].
These may be fitted by replacing $\vec e$ by $W$ for decays via
the $\omega$ and by $k(K_1) - k(K_2)$ for decays via the $\phi$.

Amplitudes for $\bar D D^*$ decays may be constructed in an analogous
fashion. For $C +1$, it is necessary to take the combination
$\bar D D^* + \bar D^* D$; for $C = -1$, the combination
$\bar D D^* - \bar D^* D$ in needed.
The vector describing $D^* \to D\pi$ is given by the 4-vector
$V^1 = (k_D - k_\pi ) -(M^2_D/M^2_{D*})(k_D + k_\pi)$.
For $D^* \to D\gamma$, the vector $\vec e$ of the $J/\Psi$ is replaced
by the corresponding vector for the photon.

Amplitudes for decay to $\gamma \chi$ are formed using
the photon direction instead of the lepton  from $J/\Psi$
decays. The vector $\vec e$ so formed then needs to be combined with
a vector constructed from $\chi _1$ decays or a tensor
for $\chi _2$ decays.

A check on programming and amplitudes is that they should be
orthogonal after integrating over all space.
If formulae for higher spins are required, or for decays of $B^*$,
the methods described by Zou and Bugg [34] and Chung [35] are useful
sources.

\section {Summary}
In Section 2, it has been shown that $X(3872)$ decay via the
$\pi \pi$ S-wave gives a poor fit to Belle data;
$J^{PC} = 1^{--}$ is not completely ruled out but would
point to the vector glueball suggested by Seth [24].
Quantum numbers $1^{+-}$, $2^{--}$ and $3^{--}$ are excluded
by the poor fits with $L = 1$ and 2.

Decays to  $\rho J/\Psi $ with
$J^{PC} = 0^{++}$, $1^{++}$ or $2^{++}$ are strong candidates;
$0^{-+}$ and $2^{-+}$ are less likely alternatives. The latter could be
eliminated if the $\pi \pi$ mass spectrum remains unchanged with a
factor 4 higher statistics. They also have characteristic angular
dependence discussed below.
Higher statistics would also discriminate for or against
$1^{--} \to [S J/\Psi ]_{L=0}$.

Suggestions will be made here on how angular momentum analysis might
proceed. Amplitudes with different $J^P$ are orthogonal, though
experimental cuts might spoil this orthogonality to some extent. The
$X(3872)$ is narrow and presumably a single state. The dependence of
amplitudes on five angles is distinctive and a separation between
different $J^P$ with forseeable statistics of 200 events looks
realistic. For $Y(3941)$, more than one $J^P$ might contribute, making
analysis more difficult.

Amplitudes for $Y(3941) \to \omega J/\Psi$ depend on the vector
$W_\alpha$, eqn. (5), which lies normal to the decay plane of the
$\omega$.
Amplitudes for $X(3872) \to \rho J/\Psi$ depend on
the vector $Q_\alpha = [k(\pi ^+) - k(\pi ^-)]_\alpha$ of eqn. (4).

Firstly, the amplitude for $J^{PC} = 1^{--} \to [S J/\Psi ]_{L=0}$
is $K.e$.
Since $e$ lies normal to the direction of the lepton from
$J/\Psi $ decay, the amplitude is a maximum with $K$
normal to that direction. The intensity depends on $\sin ^2 \alpha$,
where $\alpha$ is the angle between kaon and lepton.
Using the mass spectrum in addition, it should be
possible to discriminate for or against this possibility.

For $\rho J/\Psi$ and $\omega J/\Psi$, decay angular
distributions are distinctively different for $J^{PC} = 0^{++}$,
$1^{++}$ and $2^{++}$, as shown by Rosner [20]. Denoting by $\Theta$
the angle between the lepton from $J/\Psi$ decay and $W$ (or $Q$), the
decay angular distribution is proportional to $\sin ^2 \Theta$ for
$0^{++}$, $(1 + \cos ^2 \Theta )$ for $1^{++}$ and $(7 - \cos ^2 \Theta
)$ for $2^{++}$. The $0^{++}$ amplitude is independent of
the kaon direction, since there is no orbital angular momentum in the
production process.

For $1^{++}$, the full matrix element is proportional to the
scalar product $e.W \wedge K$ or
$e.Q \wedge K$ for decays with $L = 0$.
The amplitude is a maximum with the kaon direction perpendicular
to $W$ (or $Q$).
There is then a $\sin ^2 \beta$ dependence on the angle $\beta$
between the kaon direction and $K$ or $Q$.
Furthermore, $W\wedge K$ or $Q\wedge K$ will have a sine squared
intensity variation with respect to the lepton direction.
The overall angular dependence is much more distinctive
than that on $\Theta$ alone.
If $L = 2$ decays of $Y(3941)$ also contribute, it is essential
to constrain the analysis with the dependence of this amplitude
on $J^2$, eqn. (13); here $J$ is the 3-momentum of the $J/\Psi$
in the
rest frame of $Y$.
The decay angular distribution is still proportional to
$(1 + \cos ^2 \Theta )$.

For $J^{PC} = 2^{++}$, there are two distinctive features.
The first is the decay angular distribution
$(7 - \cos ^2 \Theta )$ , which is the same for all $2^+$
amplitudes.
The second feature is the dependence on production
through the tensor $\tau _{\alpha \beta }$ of eqn. (11).
For $Y(3941)$, the available momentum in the decay
is 282 MeV/c at the peak. The suppression of the $L=2$
amplitude by a conventional centrifugal barrier of radius
0.8 fm is then a factor 3 in amplitude, 9 in intensity.
It is therefore likely that $L = 0$ decays would dominate.
The first step is to look for this amplitude
$\tau _{\alpha _\beta }T^{\alpha \beta }$ where
$T$ is given by eqn. (10).
The intensity has a fourth order dependence on the
angle of the kaon with respect to the lepton from $J/\Psi$
decay and a second order dependence on the angle between $W$
and the lepton. This correlation is distinctive.
If such angular dependence is observed, additional $L=2$ amplitudes
can be tried one by one.
With modest statistics, it is not a good idea to fit using all
four $2^+$ amplitudes, since
experience elsewhere is that linear combinations of these
four amplitudes can usually simulate lower spins 1 and 0.
High statistics would be needed to separate
the three $L=2$ amplitudes with combined spins $s = 0$, 1 and 2
between $\omega$ and $J/\Psi$ and  this may not be of great interest.

The $J^{PC} = 0^{-+}$ amplitude, eqn. (20), may be written as the
scalar product $e.J\wedge W$.
The matrix element is a maximum when $J$ and $W$ are orthogonal
and their vector product is normal to the lepton direction.
The intensity has a sine squared dependence
on the angle between $J$ and $W$ and on the angle between their vector
product and the lepton. The amplitude is unique and angular
correlations are very distinctive, so identifying this
amplitude should be easy.

For $J^{PC} = 2^{-+}$ there are two amplitudes, making analysis more
difficult. As for $2^{++}$, a characteristic signature is the
dependence on $\tau _{\alpha \beta }$, hence a fourth order
dependence of intensity on the angles between the kaon and
$W$, $J$ or the lepton direction.
The analysis is most easily done by fitting eqns. (17-19) to the
data individually, then in combination. Again one should beware
of two $2^{-+}$ amplitudes simulating lower spins.

Now consider other decays.
Information on the $\bar D D^*$ channel is of primary
importance.
Formulae given here apply also to this channel and also decays to
$\gamma \chi$ and $S \eta _c$, with simple substitutions of
variables.
If the $X(3872)$ is a $1^{++}$ bound state, sufficient statistics
must reveal decays to $\bar D D^*$ peaking in the first 10 MeV
above threshold (in analogy with the $NN$ $^3S_1$ cross
section, which is related to the existence of the deuteron).
Evidence for $1^{++}$ D-wave decays would favour a bound-state;
their absence would favour a cusp.
The angular momentum analysis is most easily done by fitting
S and D-wave $1^{++}$ amplitudes directly to data. Decays
through the D-wave would appear through interference with the S-wave
and a distinctive dependence on $J^2$, where $J$ describes the
direction of $\bar D$ or in the rest frame of $\bar D D^*$. This
requires good mass resolution and quite high statistics.

Charmonium radial excitations with $J^{PC} = 0^{++}$ or $2^{++}$
are likely to have decays to $\bar D D$ and $\bar D_s D_s$, so
information on these channels is important.
For spin 0, production and decay are isotropic.
For spin 2, there is an intensity dependence $(3\cos ^2 \gamma - 1)^2$,
where $\gamma$ is the angle between $\bar D$ and the direction of the
recoil kaon in the rest frame of $Y$.
However, the weak process $B \to K + 2^{++}$ is first forbidden, so
the $2^{++}$ state may be suppressed there.

Searches for open channels such as $[S \eta _c]_{L=1}$
are important.
A knowledge of the rate of this process and that to $\bar D D^*$
would allow a proper calculation of the magnitude to be expected
from a cusp.

\section {Acknowledgements}
I am grateful to Andrei Sarantsev, Volodya Anisovich and  Bing
Song Zou for enlightenment on methods of tensor algebra over
a period of many years. I also wish to thank Dr. S. Solsen
for valuable comments on Belle data.

\section {Appendix}
In the $J/\Psi$ rest frame, the component of $J/\Psi$ polarisation
along the lepton axis is 0. Here, this polarisation is transformed
first to Rosner's axes.
Then the effect of the Lorentz transformation to the rest frame of
$Y(3941)$ is evaluated.

In my axes, $\vec e$ has components $(\cos R, \sin R, 0)$, where
$R$ is unknown and must be averaged from 0 to $2\pi$.
In Rosner's axes, $\vec e$ has components:
\begin {eqnarray}
\vec e &=& \left(\begin{array}{ccc}
\cos \theta & 0 & -\sin \theta  \\
0 & 1 & 0 \\
\sin \theta  & 0 & \cos \theta
\end{array}\right)
\left(\begin{array}{ccc}
\cos \phi & \sin \phi & 0 \\
-\sin \phi & \cos \phi & 0 \\
0 & 0 & 1
\end{array}\right)
\left(\begin{array}{c}
\cos R \\
\sin R \\
0
\end{array}\right)  \\
&=& \left(\begin{array}{c}
\cos \phi\cos \theta \cos R - \sin \phi \sin R \\
\sin \phi \cos \theta \cos R + \cos \phi \sin R \\
-\sin \theta \cos R
\end{array}\right) .
\end{eqnarray}
Since $e$ is a vector,
the Lorentz transformation to the rest frame of $Y(3941)$
changes the $Z$-component to $-\gamma \sin \theta \cos R$
and produces a time-component $\beta \gamma \sin \theta \cos R$.
Rotating back through angles $\phi $ and $\theta$
to axes in the $Y$ rest frame parallel to the original $z$-axis,
the result is a 4-vector
\begin {equation}
\vec e = \left(\begin{array}{c}
[1 + (\gamma - 1)\sin ^2 \theta]\cos R \\
\sin R \\
\sin \theta \cos \theta (1 - \gamma) \sin R \\
\beta \gamma \sin \theta \cos \theta \cos R
\end {array}\right).
\end {equation}
For both $Y(3941)$ and $X(3872)$, $\beta$ is small, so
the Lorentz transformation has only a small effect on $\vec e$.

\end {document}